\documentclass[11pt]{article}

\usepackage{amsmath,amsfonts,amssymb}
\usepackage{graphicx} 
\usepackage{cite}
\usepackage{url}
\usepackage{multirow}
\usepackage[margin=0.8in]{geometry}

\title{ProDiGe: PRioritization Of Disease Genes with multitask machine learning from positive and unlabeled examples}
\author{
\bf Fantine Mordelet$^{1,2,3,4}$ and Jean-Philippe Vert$^{1,2,3}$\\
$^1$Mines ParisTech, Centre for Computational Biology, Fontainebleau, F-77300, France\\
$^2$ Institut Curie, Paris, F-75248 France\\
$^3$ INSERM, U900, Paris, F-75248 France\\
$^4$ CREST, INSEE, Malakoff, F-92240 France\\
 \texttt{fantinemordelet@gmail.com}  \hspace{1cm}     \texttt{jean-philippe.vert@mines.org}
}

\begin{document}

\maketitle

\begin{abstract}
Elucidating the genetic basis of human diseases is a central goal of genetics and molecular biology. While traditional linkage analysis and modern high-throughput techniques often provide long lists of tens or hundreds of disease gene candidates, the identification of disease genes among the candidates remains time-consuming and expensive. Efficient computational methods are therefore needed to prioritize genes within the list of candidates, by exploiting the wealth of information available about the genes in various databases. Here we propose ProDiGe, a novel algorithm for Prioritization of Disease Genes. ProDiGe implements a novel machine learning strategy based on learning from positive and unlabeled examples, which allows to integrate various sources of information about the genes, to share information about known disease genes across diseases, and to perform genome-wide searches for new disease genes. Experiments on real data show that ProDiGe outperforms state-of-the-art methods for the prioritization of genes in human diseases.
\end{abstract}

\section*{Introduction}
\label{section:intro}
During the last decades, considerable efforts have been made to elucidate the genetic basis of rare and common human diseases. The discovery of so-called \emph{disease genes}, whose disruption causes congenital or acquired disease, is indeed important both towards diagnosis and towards new therapies, through the elucidation of the biological bases of diseases. Traditional approaches to discover disease genes first identify chromosomal regions likely to contain the gene of interest, e.g., by linkage analysis or study of chromosomal aberrations in DNA samples from large case-control populations. The regions identified, however, often contain tens to hundreds of candidate genes. Finding the causal gene(s) among these candidates is then an expensive and time-consuming process, which requires extensive laboratory experiments. Progresses in sequencing, microarray or proteomics technologies have also facilitated the discovery of genes whose structure or activity are modified in disease samples, on a full genome scale. However, again, these approaches routinely identify long lists of candidate disease genes among which only one or a few are truly the causative agents of the disease process, and further biological investigations are required to identify them. In both cases, it is therefore important to select the most promising genes to be further studied among the candidates, i.e., to \emph{prioritize} them from the most likely to be a disease gene to the less likely.\\

Gene prioritization is typically based on prior information we have about the genes, e.g., their biological functions, patterns of expression in different conditions, or interactions with other genes, and follows a ``guilt-by-association'' strategy: the most promising candidates genes are those which share similarity with the disease of interest, or with other genes known to be associated to the disease. The availability of complete genome sequences and the wealth of large-scale biological data sets now provide an unprecedented opportunity to speed up the gene hunting process \cite{Giallourakis2005Disease}. Integrating a variety of heterogeneous information stored in various databases and in the literature to obtain a good final ranking of hundreds of candidate genes is, however, a difficult task for human experts. Unsurprisingly many computational approaches have been proposed to perform this task automatically via statistical and data mining approaches. While some previous works attempt to identify promising candidate genes without prior knowledge of any other disease gene, e.g., by matching the functional annotations of candidate genes to the disease or phenotype under investigation \cite{Perez-Iratxeta2002Association, Turner2003POCUS,Tiffin2005Integration}, many successful approaches assume that some disease genes are already known and try to detect candidate genes which share similarities with known disease genes for the phenotype under investigation \cite{Freudenberg2002similarity-based,Aerts2006Gene,Bie2007Kernel-based,Linghu2009Genome-wide,Hwang2010Heterogeneous,Yu2010L2-norm} or for related phenotypes \cite{Freudenberg2002similarity-based,Ala2008Prediction,Wu2008Network-based,Kohler2008Walking,Vanunu2010Associating,Hwang2010Heterogeneous}. These methods vary in the algorithm they implement and in the data they use to perform gene prioritization. For example, Endeavour and related work \cite{Aerts2006Gene,Bie2007Kernel-based,Yu2010L2-norm} use state-of-the-art machine learning techniques to integrate heterogeneous information and rank the candidate genes by decreasing similarity to known disease genes, while PRINCE \cite{Vanunu2010Associating} uses label propagation over a protein-protein interaction (PPI) network and is able to borrow information from known disease genes of related diseases to find new disease genes. We refer the reader to \cite{Tranchevent2010guide} for a recent review of gene prioritization tools available on the web.\\

Here we propose ProDiGe, a new method for prioritization of disease genes based on the guilt-by-association concept. ProDiGe assumes that a set of gene-disease associations is already known to infer new ones, and brings three main novelties compared to existing methods. First, ProDiGe implements a novel machine learning paradigm to score candidate genes. While existing methods like those of \cite{Aerts2006Gene,Bie2007Kernel-based,Yu2010L2-norm} score independently the different candidate genes in terms of similarity to known disease genes, ProDiGe exploits the relative similarity of both known and candidate disease genes to jointly score and rank all candidates. This is done by formulating the disease gene prioritization problem as an instance of the problem known as \emph{learning from positive and unlabeled examples} (PU learning) in the machine learning community, which is known to be a powerful paradigm when a set of candidates has to be ranked in terms of similarity to a set of positive data \cite{Liu2002Partially,Denis2005Learning,Mordelet2010bagging}. Second, in order to rank candidate genes for a disease of interest, ProDiGe borrows information not only from genes known to be associated to the disease, but also from genes known to play a role in diseases or phenotypes related to the disease of interest. This again differs from \cite{Aerts2006Gene,Bie2007Kernel-based,Yu2010L2-norm} which treat diseases independently from each other. It allows us, in particular, to rank genes even for \emph{orphan diseases}, with no known gene, by relying only on known disease genes of related diseases. In the machine learning jargon, we implement a \emph{multi-task} strategy to share information between different diseases \cite{Evgeniou2005Learning,Jacob2008Efficient,Jacob2008Protein}, and weight the sharing of information by the phenotypic similarity of diseases. Third, ProDiGe performs heterogeneous data integration to combine a variety of information about the genes in the scoring function, including sequence features, expression levels in different conditions, PPI interactions or presence in the scientific literature. We use the powerful framework of \emph{kernel methods} for data integration \cite{Pavlidis2002Learning,Schoelkopf2004Kernel,Lanckriet2004statistical}, akin to the work of \cite{Aerts2006Gene,Bie2007Kernel-based,Yu2010L2-norm}. This differs from approaches like that of \cite{Vanunu2010Associating}, which are limited to scoring over a gene or protein network.\\

We test ProDiGe on real data extracted from the OMIM database \cite{McKusick2007Mendelian}. It is able to rank the correct disease gene in the top 5\% of the candidate genes for 69\% of the diseases with at least one other known causal gene, and for 67\% of the diseases when no other disease genes is known, outperforming state-of-the-art methods like Endeavour and PRINCE. \\

\section*{Results}
\label{section:results}

\subsection*{Gene prioritization without sharing of information across diseases}
\label{sec:localresults}
We first assess the ability of ProDiGe to retrieve new disease genes for diseases with already a few known disease genes, without sharing information across different diseases. As a gold standard we extracted all known disease-gene associations from the OMIM database \cite{McKusick2007Mendelian}, and we borrowed from \cite{Bie2007Kernel-based} nine sources of information about the genes, including expression profiles in various experiments, functional annotations, known protein-protein interactions (PPI), transcriptional motifs, protein domain activity and literature data. Each source of information was encoded in a kernel functions, which assesses pairwise similarities between each pair of genes according to each source of information. We compare two ways to perform data integration: first by simply averaging the nine kernel functions, and second by letting ProDiGe optimize itself the relative contribution of each source of information when the model is estimated, through a multiple kernel learning (MKL) approach. We compare both variants with the best model of \cite{Yu2010L2-norm}, namely, the MKL1Class model which differs from ProDiGe in this case only in the machine learning paradigm implemented: while ProDiGe learns a model from positive and unlabeled examples, MKL1class learns it only from positive examples. We tested these three algorithm in a leave-one-out cross-validation (LOOCV) setting. In short, for each disease, each known disease gene was removed in turn, a model was trained on using the remaining disease genes as positive examples, and all 19540 genes in our database were ranked; we then recorded the rank of the positive gene that was removed in this list. We focused on the 285 diseases in our dataset having at least 2 known disease genes, because all three methods require at least one known disease gene for training, and for the purpose of LOOCV we need in addition one known disease gene removed from the training set.\\

Figure 1 
presents the cumulative distribution function (CDF) of the rank of the left-out positive gene, i.e., the number of genes that were ranked in the top $k$ genes of the list as a function of $k$, for each method. Note that the rank is always between $1$ (best prediction) and $19540-|P|$, where $|P|$ is the number of genes known to be associated to the disease of interest. The right panel zooms on the beginning of this curve which corresponds to the distribution of small values of the rank. We see clearly that both ProDiGe variants outperform MKL1class in the sense that they consistently recover the hidden positive gene at a better rank in the list. A Wilcoxon signed rank test confirms these visual conclusions at $5\%$ level with P-values $6.1e^{-29}$ and $8.8e^{-28}$, respectively, for the average and MKL variants of ProDiGe. This illustrates the benefits of formulating the gene ranking problem as a PU learning problem, and not as a 1-class learning one, since apart from this formulation both MKL1Class and ProDiGe1 use very similar learning engines, based on SVM and MKL.\\

Both ProDiGe1 variants recover roughly one third of correct gene-disease associations in the top $20$ genes among almost $19540$, i.e., in the top $0.1\%$. However, we found no significant difference between the mean and MKL variants of ProDiGe in this setting (P-value=0.619). This means that in this case, assigning equal weights to all data sources works as well as trying to optimize these weights by MKL. Supported by this result and by the fact that MKL is much more time-consuming than a SVM with the mean kernel, we decided to restrict our experiments to the mean kernel in the following experiments.\\

\begin{figure}[!ht]
\begin{center}
\includegraphics[width=\textwidth]{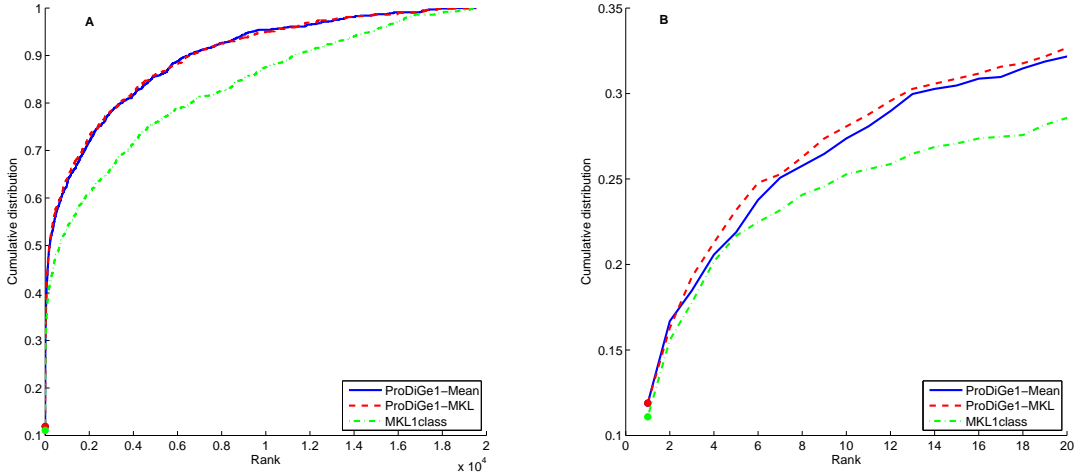}
\end{center}
\caption{
{\bf Cumulative distribution function of the rank for local methods, in the LOOCV experiment.}  ProDiGe1-Mean and ProDiGe1-MKL refer to the ProDiGe1 variant combined with the mean kernel or the MKL strategy to fuse heterogeneous gene information. Panel (A) Global curve, Panel (B) Zoom on the beginning of the curve.
}
\label{fig:CDFLocal}
\end{figure}

\subsection*{Gene prioritization with information sharing across diseases}
\label{sec:globalresults}
In a second run of experiments, we assess the performance of ProDiGe when it is allowed to share informations across diseases. We tested three variants of ProDiGe, as explained in Material and Methods: ProDiGe2, which uniformly shares information across all diseases without using particular informations about the diseases, ProDiGe3, which weights the sharing of informations across diseases by a phenotypic similarity between the diseases, and ProDiGe4, a variant of ProDiGe3 which additionally controls the sharing of information between diseases that would have very similar phenotypic description but which remain different diseases. All variants are based on the same methodological backbone, namely, the use of a multitask learning strategy, and only differ in a function used to control the sharing of information. We limit ourselves to the 1873 diseases in the disease-gene association dataset which were also in the phenotypic similarity matrix that we used. This corresponds to a total of 2544 associations between these diseases and 1698 genes. We compare these variants  to PRINCE \cite{Vanunu2010Associating}, a method recently proposed to rank genes by sharing information across diseases through label propagation on a PPI network.\\

Figure 2 
shows the CDF curves for the four methods. Comparing areas under the global curve, i.e., the average rank of the left-out disease gene in LOOCV, the four methods can be ranked in the following order: ProDiGe4 (1682) $>$ ProDiGe3 (1817) $>$ ProDiGe2 (2246) $>$ PRINCE (3065). The fact that ProDiGe3 and ProDiGe4 outperform ProDiGe2 confirms the benefits of exploiting prior knowledge we have about the disease phenotypes to weight the sharing of information across diseases, instead of following a generic strategy for multitask learning. The fact that ProDiGe4 outperforms ProDiGe3 is not surprising and illustrates the fact that the diseases are not fully characterized by the phenotypic description we use. Zooming to the beginning of the curves (right picture), we see that the relative order between the methods is conserved except for PRINCE which outperforms ProDiGe2 in that case. In fact, ProDiGe2 has a very low performance compared to all other methods for low ranks, confirming that the generic multitask strategy should not be pursued in practice if phenotypic information is available.\\

The fact that ProDiGe3 and ProDiGe4 outperform PRINCE for all rank values confirm the competitiveness of our approach. On the other hand, the comparison with PRINCE is not completely fair since ProDiGe exploits a variety of data sources about the genes, while PRINCE only uses a PPI network. In order to clarify whether the improvement of ProDiGe over PRINCE is due to a larger amount of data used, to the learning algorithm, or to both, we ran ProDiGe3 with only the kernel derived from the PPI network which we call ProDiGe-PPI in Figure 2. 
In that case, both ProDiGe and PRINCE use exactly the same information to rank genes. We see on the left picture that this variant is overall comparable to PRINCE (no significant difference between PRINCE and ProDiGe-PPI with a Wilcoxon paired signed rank test), confirming that the main benefit of ProDiGe over PRINCE comes from data integration. Interestingly though, at the beginning of the curve (right picture), ProDiGe-PPI is far above PRINCE, and even behaves comparably to the best method ProDiGe4. Since ProDiGe-PPI and PRINCE use exactly the same input data, this means that the better performance of ProDiGe-PPI for low ranks comes from the learning method based on PU learning with SVM, as opposed to label propagation over the PPI network.\\

\begin{figure}[!ht]
\begin{center}
\includegraphics[width=\textwidth]{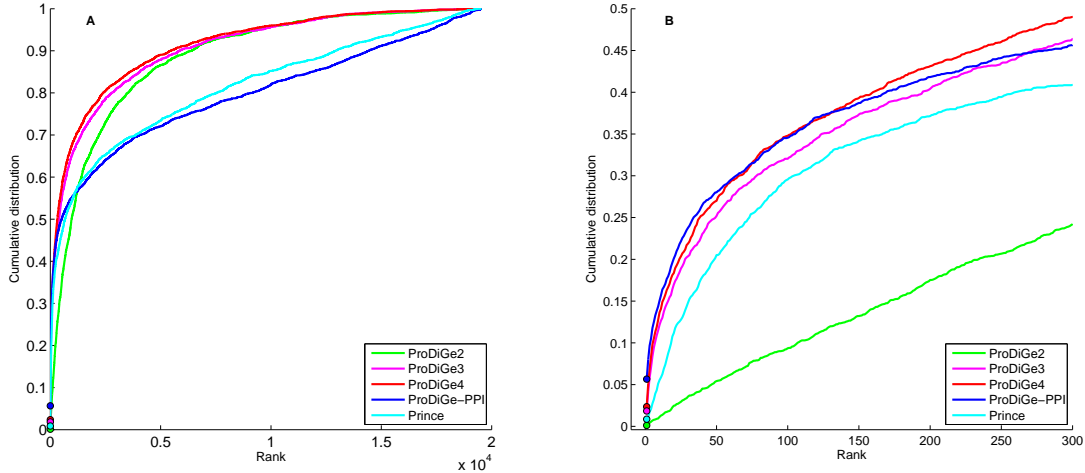}
\end{center}
\caption{
{\bf Cumulative distribution function of ranks in the LOOCV experiments, for global approaches 
.} ProDiGe2, 3, 4 refer to the three variants of ProDiGe which share information, while ProDiGe-PPI refers to ProDiGe3 trained only the PPI network data. Panel (A) Global curve. Panel (B) Zoom on the beginning of the curve.
}
\label{fig:cdfMultitask}
\end{figure}

To better visualize the differences between the different variants of ProDiGe, the scatter plots in Figure 3 
compare directly the ranks obtained by the different variants for each of the 2544 left-out associations. Note that smaller ranks are better than large ones, since the goal is to be ranked as close as possible to the top of the list. On the left panel, we compare ProDiGe3 to ProDiGe4. We see that many points are below the diagonal, meaning that adding a Dirac kernel to the Phenotype kernel (ProDiGe4) generally improves the performance as compared to using a Phenotype kernel (ProDiGe3) alone. On the right panel, the ProDiGe2 is compared to the ProDiGe3. We see that the points are more concentrated above the diagonal, but with large variability on both sides of the diagonal. This indicates a clear advantage in favor of the Phenotype kernel compared to the generic Multitask kernel, although the differences are quite fluctuant.\\

\begin{figure}[!ht]
\begin{center}
\includegraphics[width=\textwidth]{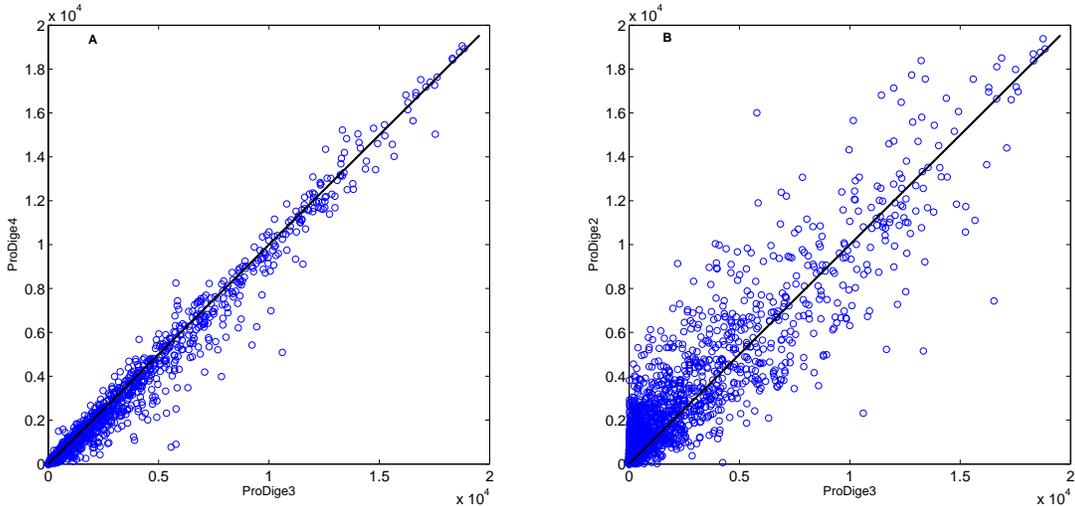}
\end{center}
\caption{
{\bf Comparison of rank measures between different variants of ProDiGe.} Each point represent a disease-gene association. We plot the rank they obtain from the different methods when they are left out in the LOOCV procedure. Small rank are therefore better than large ranks.
}
\label{fig:scatter}
\end{figure}

\subsection*{Is sharing information across diseases beneficial?}
In order to check whether sharing information across diseases is beneficial, we restrict ourselves to diseases with phenotypic informations and with at least two known associated genes in the OMIM database. This way, we are able to share information across diseases and, at the same time, to run methods that do not share information because we ensure that there is at least one training gene in the LOOCV procedure. This leaves us with 265 diseases, corresponding to 936 associations.\\

Figure 4 
shows the CDF curves of the rank for the various methods on these data, including the two methods MKL1class and ProDiGe1 (with the mean kernel for data integration), which do not share information across diseases, and ProDiGe 2, 3, 4 and PRINCE, which do share information. Interestingly, we observe different retrieval behaviors on these curves, depending on the part of the curve we are interested in. On the one hand, if we look at the curves globally, ProDiGe 4 and 3 perform very well, having high area under the CDF curve, i.e., a low average rank (respectively 1529 and 1770). PRINCE and MKL1class have the worse average ranks (respectively 3220 and 3351). A systematic test of differences between the methods, using a Wilcoxon paired signed rank test over the ranks for each pair of methods, is summarized in Figure 5 
. In this picture, an arrow indicates that a method is significantly better than another at level $5\%$. This confirms that ProDiGe 4 is the best method, significantly better than all other ones except ProDiGe 1. Three variants of ProDiGe are significantly better than PRINCE and MKL1Class.\\
 
On the other hand, in the context of gene prioritization, it is useful to focus on the beginning of the curve and not on the full CDF curves. Indeed, only the top of the list is likely to deserve any serious biological investigation. Therefore we present a zoom of the CDF curve in panel (B) of Figure 4 
. We see there that the local methods ProDiGe1 and MKL1class present a sharper increase at the beginning of the curve than the global methods, meaning that they yield more often truly disease genes near the very top of the list than other methods. Additionally, we observe that ProDiGe1 is in fact the best method when we focus on the proportion of disease genes correctly identified in up to the top 350 among 19540, i.e., in up to the top 1.8\% of the list. These results are further confirmed by the quantitative values in Table \ref{tab:precision}, which show the recall (i.e., CDF value) as a function of the rank. ProDiGe 1, which does not share information across diseases, is the best when we only focus at the very top of the list (up to the top 1.8\%), while ProDiGe 4, which shares information, is then the best method when we go deeper in the list. \\

\begin{table}[ht!]
\begin{center}
\begin{tabular}{l|c|c|c|c|c}
 & top 1 & top 10 & top $1\%$ & top $5\%$ & top $10\%$ \\ 
 \hline
MKL1class & 11.5 & 25.3 & 41.1 & 52.8 & 59.9 \\ 
ProDiGe1 & \textbf{12.3} & \textbf{27.8} & \textbf{49.2} & 61.9 & 71.2 \\ 
ProDiGe2 & 0.1 & 0.7 & 17.8 & 51.2  & 66.9 \\ 
ProDiGe3 & 1.9 & 11.4 & 38.6 & 64.0 & 74.2 \\ 
ProDiGe4 & 3.1 & 14.6 & 43.4 & \textbf{68.9} & \textbf{78.4} \\ 
PRINCE & 1.5 & 6.8 & 37.3 & 57.1 & 65.4
\end{tabular}
\end{center}
\caption{{\bf Recall of different methods at different rank levels, for diseases with at least one known disease gene.}  The recall at rank level $k$ is the percentage of disease genes that were correctly ranked in the top $k$ candidate genes in the LOOCV procedure, where the number of candidate genes is near $19540$. Top 1 and top 10 (first two columns) correspond respectively to the recall at the first and first ten genes among 19540, while top X\% (last three columns) refer to the recall at the first X\% genes among 19540.}
\label{tab:precision}
\end{table}

At this point it is interesting to question what position in the list we are interested in. In classical applications where we start from a short list of, say, 100 candidates, then being in the top $5\%$ of the list means that the correct gene is ranked in the top 5 among the 100 candidates, while the top $1\%$ corresponds to the first of the list (see the last 3 columns of table \ref{tab:precision}). If we only focus on the first gene of a short list of 100 candidates, then ProDiGe1 is the best method, with almost half of the genes ($49.2\%$) found in the first position, followed by ProDiGe4 ($43.4\%)$ and MKL1class ($41.1\%$). As soon as we accept to look further than the first place only, ProDiGe 4 is the best method, with $68.9\%$ of disease genes in the top 5 of a list of 100 candidates, for example. On the other hand, if we consider a scenario where we start from no short list of candidates, and directly wish to predict disease genes among the 19540 human genes, then only the few top genes in the list are interesting (see the first 2 columns of table \ref{tab:precision}). In that case, the methods that do not share information are clearly preferable, with $27.8\%$ (resp $25.3\%$) of genes correctly found in the top 10 among 19540 for ProDiGe 1 (resp. MKL1class).\\

In summary, sharing information is not beneficial if we are interested only in the very top of the list, typically the top 10 among 19540 candidates. This setting is however very challenging, where even the best method ProDiGe1 only finds $12.3\%$ of all disease genes. As soon as we are interested in more than the top $2\%$ of the list, which is a reasonable level when we start from a short list of a few tens or hundreds of candidate genes, sharing information across diseases becomes interesting. In all cases, some variant of ProDiGe outperforms existing methods. In particular ProDiGe4, which shares information using phenotypic information across diseases while keeping different diseases distinct, is the best way to share information.\\

 \begin{figure}[!ht]
\begin{center}
\includegraphics[width=\textwidth]{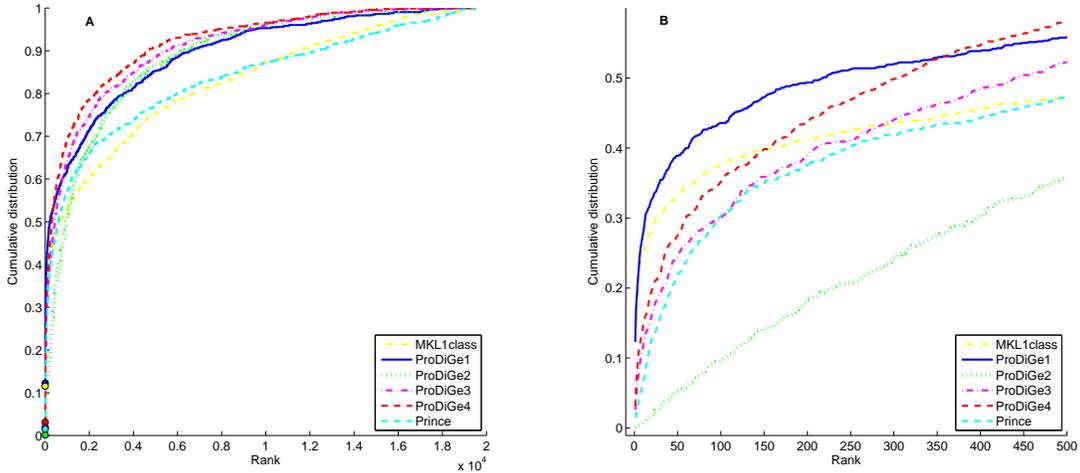}
\end{center}
\caption{
{\bf Cumulative distribution function of ranks for local and multitask approaches.} (A) Global curve. (B) Zoom on the beginning of the curve.
}
\label{fig:cdfMulti}
\end{figure}

  \begin{figure}[!ht]
\begin{center}
\includegraphics[width=0.5\textwidth]{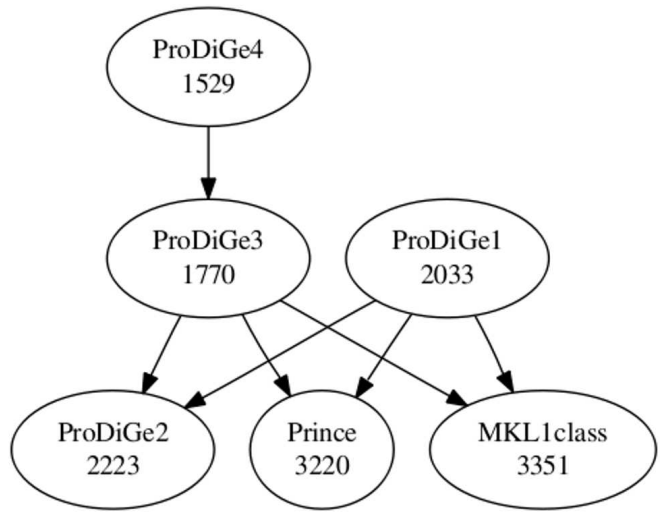}
\end{center}
\caption{
{\bf Wilcoxon paired signed rank tests for significant rank difference between all methods.} ProDiGe1 and MKL1class are the only local approaches, which do not share information across diseases. The number in each ellipse is the average rank obtained by the method in the LOOCV procedure. An arrow indicates that a method is significantly better than another.
}
\label{fig:wilcoxon}
\end{figure}

\subsection*{Predicting causal genes for orphan diseases}
Finally, we investigate the capacity of the different gene prioritization methods to identify disease genes for orphan diseases, i.e., diseases with no known causative gene yet. ProDiGe1 and MKL1class, which treat diseases independently from each other and require known disease genes to find new ones, can not be used in this setting. Methods that share information across diseases, i.e., ProDiGe2, 3, 4 and PRINCE, can be tested in this context, since they may be able to discover causative genes for a given orphan diseases by learning from causative genes of other diseases. In fact, ProDiGe3 and ProDiGe4 boil down to the same method in this context, because the contribution of the Dirac kernel in (\ref{eq:kernelPD}) vanishes when no known disease gene for a disease of interest is available during training. We summarize them by the acronym ProDiGe3-4 below.\\

To simulate this setting, we start from the 1608 diseases with only one known disease gene in OMIM and phenotypic information, resulting in 1608 disease-gene associations involving 1182 genes. For each disease in turn, the associated gene is removed from the training set, a scoring function is learned from the associations involving other diseases, and the removed causal gene is ranked for the disease of interest. We compute the rank of the true disease gene, and repeat this operation for each disease in turn. Figure 6 
and Table \ref{tab:precisionDN} show the performance of the different global methods in this setting. Interestingly, they are very similar to the results obtained in the multitask setting (Figure 2 
and Table  \ref{tab:precision}), both in relative order of the methods and in their absolute performance. Overall, ProDiGe3-4 performs best, retrieving the true causal gene in the top 10 genes of the list $13.1\%$ of times, and in the top $5\%$ of candidate genes $66.9\%$ of times. This is only slightly worse than the performance reached for diseases with known disease genes (respectively $14.6\%$ and $68.9\%$), highlighting the promising ability of global approaches to deorphanize diseases.\\

 \begin{figure}[!ht]
\begin{center}
\includegraphics[width=\textwidth]{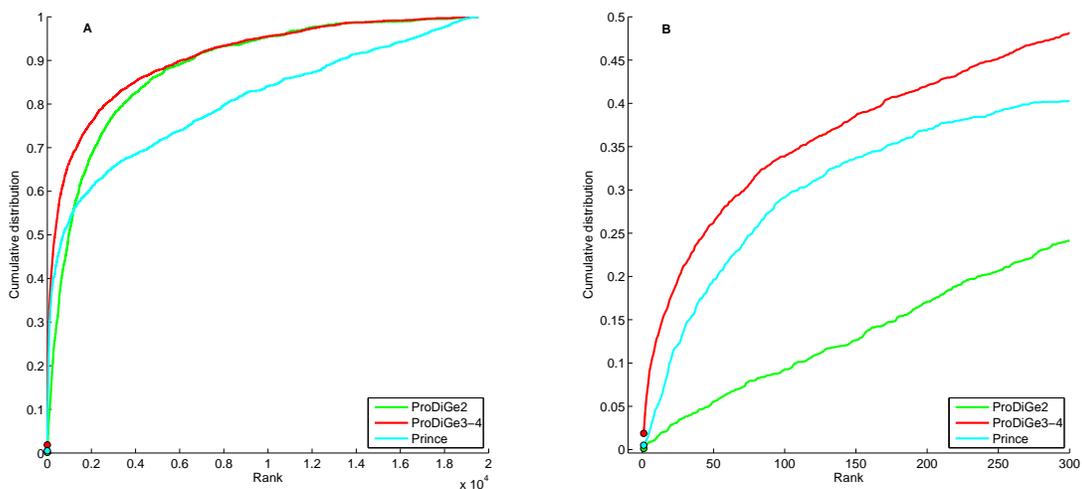}
\end{center}
\caption{
{\bf Cumulative distribution function of ranks for prioritization of causal genes for orphan diseases.} Panel (A) Global curve. Panel (B) Zoom on the beginning of the curve.
}
\label{fig:cdfMultitaskOrphan}
\end{figure}

\begin{table}[ht!]
\begin{center}
\begin{tabular}{l|c|c|c|c|c}
& top 1 & top 10 & top $1\%$ & top $5\%$ & top $10\%$ \\  
 \hline
ProDiGe2 & 0.1 & 1.4 & 16.8 & 50.4 & 68.1 \\ 
ProDiGe3-4 & \textbf{1.9} & \textbf{13.1} & \textbf{42.7} & \textbf{66.9} & \textbf{76.1} \\ 
PRINCE & 0.5 & 4.8 & 36.9 & 52.9 & 60.6
\end{tabular}
\end{center}
\caption{ {\bf Recall of different methods at different rank levels, for orphan diseases.} 
In this case, since the disease has no known causal genes, only the causal genes of other diseases intervene in the learning, meaning that ProDiGe3 and 4 are equivalent approaches.}
\label{tab:precisionDN}
\end{table}

\subsection*{Validation on selected diseases}
To further validate ProDiGe, we used the whole training set to prioritize the unlabeled genes for a few particular diseases with ProDiGe4. We completed the training set with a list of genes collected through the use of Ingenuity Pathways Analysis (IPA, Ingenuity® Systems). In Table \ref{tab:ingenuity}, we report the results of this validation for a first set of diseases having a training set of positive genes of reasonable size (more than 11 genes). These diseases are in the same order: prostate cancer [MIM 176807], colorectal cancer [MIM 114500], diabetes mellitus [MIM 125853], Alzheimer [MIM 104300], gastric cancer [MIM 137215], leukemia acute myeloid [MIM 601626], breast cancer [MIM 114480], schizophrenia [MIM 181500]. The columns report successively the disease name, the MIM id of the disease, the size of the training set, the size of the intersection between the training set and the Ingenuity list, the estimated precision and recall of the top 100 genes in the prioritized list and the p-value of a hypergeometric test. The precision is estimated as the fraction of the top 100 genes that are also in the IPA list while recall is the fraction of the IPA list that intersects the top 100 genes of the prioritized list. Of course, the true precision value is unknown and the value we report underestimates the true value. The hypergeometric test allows to test for the enrichment of the top 100 genes of our prioritized list in genes known to be associated to the disease, which were not in the training set (namely genes previously extracted from IPA). We can see that precision is good, except for schizophrenia, gastric cancer and leukemia. Recall on the other hand is not very high but the values are limited by the large size of IPA lists. All tests are significant at $5\%$ level.\\

\begin{table}[ht!]
\begin{center}
\begin{tabular}{l|c|c|c|c|c|c}
Disease name & MIM Id  & Training set & Training $\bigcap$ IPA  & Precision (\%) & Recall (\%) & P-value\\ 
\hline
Prostate cancer & 176807  & 12 & 12 & 41 & 7.5 & $5.3e^{-40}$\\ 
Colorectal cancer & 114500  & 17 & 17 & 51 & 5.7 & $7.3e^{-44}$\\ 
Diabetes mellitus & 125853  & 26 & 22 & 21 & 1.4 & $2.1e^{-06}$\\ 
Alzheimer & 104300  & 11 & 10 & 23 & 2.3 & $3.8e^{-11}$\\ 
Gastric cancer & 137215  & 12 & 12 & 16 & 7.1 & $9.3e^{-16}$\\ 
Leukemia acute myeloid & 601626  & 17 & 16 & 13 & 10.0 & $2.8e^{-15}$\\ 
Breast cancer & 114480  & 19 & 16 & 33 & 3.7 & $6.4e^{-22}$\\ 
Schizophrenia & 181500  & 17 & 11 & 6 & 3.2 & $4.5e^{-05}$
 \end{tabular}
 \end{center}
\caption{{\bf Prioritization with ProDiGe4 for 8 diseases with a large training set of known genes.} 
 The results were validated by comparing our top 100 genes with a list of genes related to the disease, extracted from Ingenuity database.}
\label{tab:ingenuity}
\end{table}

We then did the same for 8 diseases with only 2 known genes in our training set: glaucoma [MIM 606657], Creutzfeld-Jacob [MIM 123400], hyperparathyroidism [MIM 145000], psoriasis [MIM 177900], glioblastoma [MIM 137800], cystic fibrosis [MIM 219700], pancreatic carcinoma [MIM 260350], thalassemia [MIM 604131]. Results are given in Table \ref{tab:ingenuity2}. As expected, precision is much smaller for these diseases. However, we see that sharing information across diseases still allows to retrieve new disease genes for diseases where the training set is very small.\\

\begin{table}[ht!]
\begin{center}
\begin{tabular}{l|c|c|c|c|c}
Disease name & MIM Id  &  Training $\bigcap$ IPA  & Precision (\%) & Recall (\%) & P-value\\ 
\hline
Glaucoma & 606657  & 2 & 8 & 12.5 & $2.0e^{-11}$\\ 
Creutzfeld-Jacob & 123400  & 2 & 2 & 40.0 & $1.3e^{-06}$\\ 
Hyperparathyroidism & 145000 & 2 & 3 & 18.7 & $1.1e^{-06}$\\ 
Psoriasis & 177900  & 2 & 4 & 6.0 & $1.8e^{-05}$\\ 
Glioblastoma & 137800  & 2 & 16 & 10.7 & $8.4e^{-19}$\\ 
Cystic fibrosis & 219700  & 2 & 5 & 10.6 & $9.3e^{-08}$\\ 
Pancreatic carcinoma & 260350  & 1 & 8 & 9.6 & $2.3e^{-10}$\\ 
Thalassemia & 604131  & 0 & 2 & 25.0 & $2.6e^{-06}$
 \end{tabular}
 \end{center}
\caption{{\bf Prioritization with ProDiGe4 for 8 diseases with only 2 known genes.} 
The results were validated by comparing our top 100 genes with a list of genes related to the disease, extracted from Ingenuity database.}
\label{tab:ingenuity2}
\end{table}

Further validation include Table \ref{tab:genelist} which reports the top ten genes of the prioritized list for prostate cancer, colorectal cancer, diabetes mellitus, Alzheimer, gastric cancer, leukemia acute myeloid, breast cancer, schizophrenia. These lists were analyzed with GeneValorization \cite{Brancotte2011Gene}, a text-mining tool for automatic bibliography search.

\begin{table}[ht!]
\begin{center}
  \begin{tabular}{|l|c|c|l|l|c|c|}
\hline
\multicolumn{3}{|c|}{Prostate cancer } && \multicolumn{3}{|c|}{Gastric cancer }\\
\hline
CDKN2A(1029)&210&1&&EGFR(1956)&853&1\\
\hline
AKT1(207)&1058&1&&AKT1(207)&272&0\\
\hline
IGF1R(3480)&152&1&&EXT1(2131)&4&0\\
\hline
MSX1(4487)&5&0&&FAS(355)&180&0\\
\hline
PAX3(5077)&2&0&&LRP5(4041)&8&0\\
\hline
CCND1(595)&372&1&&MSX1(4487)&3&0\\
\hline
BRAF(673)&22&1&&CCND1(595)&250&1\\
\hline
TP53(7157)&1378&1&&BRAF(673)&32&1\\
\hline
WFS1(7466)&0&0&&TP53(7157)&1593&1\\
\hline
WT1(7490)&37&1&&WFS1(7466)&0&0\\
\hline
\multicolumn{3}{|c|}{Colorectal cancer }&& \multicolumn{3}{|c|}{Leukemia acute myeloid}\\
\hline
CDKN2A(1029)&415&1&&AKT1(207)&233&0\\
\hline
EXT1(2131)&14&0&&FAS(355)&136&0\\
\hline
IGF1R(3480)&86&1&&KRAS(3845)&457&1\\
\hline
SMAD4(4089)&211&1&&LYN(4067)&26&0\\
\hline
MLH1(4292)&4064&1&&MYC(4609)&381&0\\
\hline
PDGFRA(5156)&19&1&&RAF1(5894)&30&1\\
\hline
PDGFRB(5159)&45&1&&STAT3(6774)&95&0\\
\hline
BRAF(673)&430&1&&STK11(6794)&2&0\\
\hline
WFS1(7466)&0&1&&BTK(695)&6&0\\
\hline
WT1(7490)&15&0&&TP53(7157)&474&1\\
\hline
\multicolumn{3}{|c|}{Diabetes mellitus }&& \multicolumn{3}{|c|}{Breast cancer }\\
\hline
COL1A1(1277)&4&0&&CDKN2A(1029)&572&1\\
\hline
COL2A1(1280)&6&0&&COL2A1(1280)&9&0\\
\hline
CYP3A5(1577)&5&0&&COL3A1(1281)&1&0\\
\hline
EXT1(2131)&20&1&&EXT1(2131)&22&0\\
\hline
GHR(2690)&49&0&&LRP5(4041)&51&0\\
\hline
ABCC6(368)&43&0&&MSX1(4487)&10&0\\
\hline
LEP(3952)&754&1&&PAX3(5077)&6&0\\
\hline
LRP5(4041)&58&0&&PITX2(5308)&310&1\\
\hline
CACNA1S(779)&4&0&&BRAF(673)&37&1\\
\hline
ADIPOQ(9370)&1635&1&&WFS1(7466)&4&0\\
\hline
\multicolumn{3}{|c|}{Alzheimer} && \multicolumn{3}{|c|}{Schizophrenia }\\
\hline
COL2A1(1280)&0&0&&COL1A1(1277)&0&0\\
\hline
CYP1B1(1545)&0&0&&COL2A1(1280)&0&0\\
\hline
EXT1(2131)&4&1&&ATN1(1822)&40&0\\
\hline
ALDH3A2(224)&4&0&&EXT1(2131)&20&0\\
\hline
APOE(348)&4143&1&&FGFR3(2261)&78&0\\
\hline
ABCC6(368)&10&0&&GJB1(2705)&0&0\\
\hline
LRP5(4041)&3&0&&ABCC6(368)&7&0\\
\hline
MAOA(4128)&5&1&&LRP5(4041)&4&0\\
\hline
PSEN2(5664)&635&1&&PARK2(5071)&1&0\\
\hline
WFS1(7466)&1&0&&WFS1(7466)&5&0\\
\hline
\end{tabular}
\end{center}
\caption{{\bf The top ten genes for 8 diseases with a reasonable training set are scanned.} 
These diseases are in order: prostate cancer [MIM 176807], colorectal cancer [MIM 114500], diabetes mellitus [MIM 125853], Alzheimer [MIM 104300], gastric cancer [MIM 137215], leukemia acute myeloid [MIM 601626], breast cancer [MIM 114480], schizophrenia [MIM 181500]. Using GeneValorization, we counted the number of publication hits in NCBI which are found to be relevant to a query disease and a query gene. At last, the third column indicates whether the gene belongs to the list extracted from the Ingenuity Pathways Analysis tool.
}
\label{tab:genelist}
\end{table}

\section*{Discussion}
\label{section:discussion}

We have introduced ProDiGe, a new set of methods for disease gene prioritization. ProDiGe integrates heterogeneous information about the genes in a unified PU learning strategy, and is able to share information across different diseases if wanted. We have proposed in particular two flavours for disease gene ranking: ProDiGe1, which learns new causal genes for each disease separately, based on already known causal genes for each disease, and ProDiGe4, which additionally transfers information about known disease genes across different diseases, weighting information sharing by disease phenotypic similarity. We have demonstrated the efficiency of both variants on real data from the OMIM database where they outperform Endeavour and PRINCE, two state-of-the-art gene prioritization methods.\\

A particularity of ProDiGe is the possibility to encode prior knowledge on disease relatedness through the disease kernel. While a Dirac kernel prevents sharing of information across diseases, we tested different variants to share information including a generic multitask kernel and kernels taking into account the phenotypic similarity between diseases. We demonstrated the relevance of using the phenotypic similarity, compared to the generic multitask kernel, and have enhanced it by the addition of a Dirac kernel. Given the influence of the disease kernel on the final performance of the method, we believe that there is still much room for improvement in the design of the prior, using the general ProDiGe framework. We note in particular that if other descriptors were available for phenotypes, one could also integrate these data and the prior they induce on task relatedness in the disease kernel.\\

A important question in practice is to choose between the two variants. We have seen that ProDiGe1 has higher recall in the top 1 or $2\%$ of the list, while ProDiGe4 is better after. Hence a first criterion to chose among them is the rank level that we are ready to investigate. In addition, one could think that ProDiGe1, which can not be used for orphan disease, is more generally handicapped compared to ProDiGe4 when the number of known disease genes is small, while it is in a better situation when many genes are already known. Indeed, if enough causal genes are known for a given disease, there is intuitively no need to borrow information from other diseases, which may mislead the prediction. This dependency of the relative performance of a local and a global approach on the number of training samples has previously been observed in other contexts \cite{Jacob2008Protein} where a global approach was shown to bring tangible improvements over a local one when the number of positive examples was low. We have however checked for the presence of such an effect, and found that it could not be brought to light, as illustrated in Figure 7 
which plots the mean and standard deviation of the rank of the left-out gene in LOOCV as a function of the number of known genes of the disease during training. We observe no trend indicating that the performance increases with the number of training genes, and no different behaviour between the local and multitask approaches, as long as at least one disease gene is known. This surprising finding, which is coherent with the observation that there is no big difference in performance for orphan and non-orphan diseases, suggests that the number of known disease genes in not a relevant criterion to choose between the local and multitask version of ProDiGe. Instead, we suggest in practice to use the local version ProDiGe 1 if we are interested only in genes ranked in the very top of the candidate gene lists (below the top $1\%$), and ProDiGe 4 as soon as we can afford going deeper in the list.\\

\begin{figure}[!ht]
\begin{center}
\includegraphics[width=\textwidth]{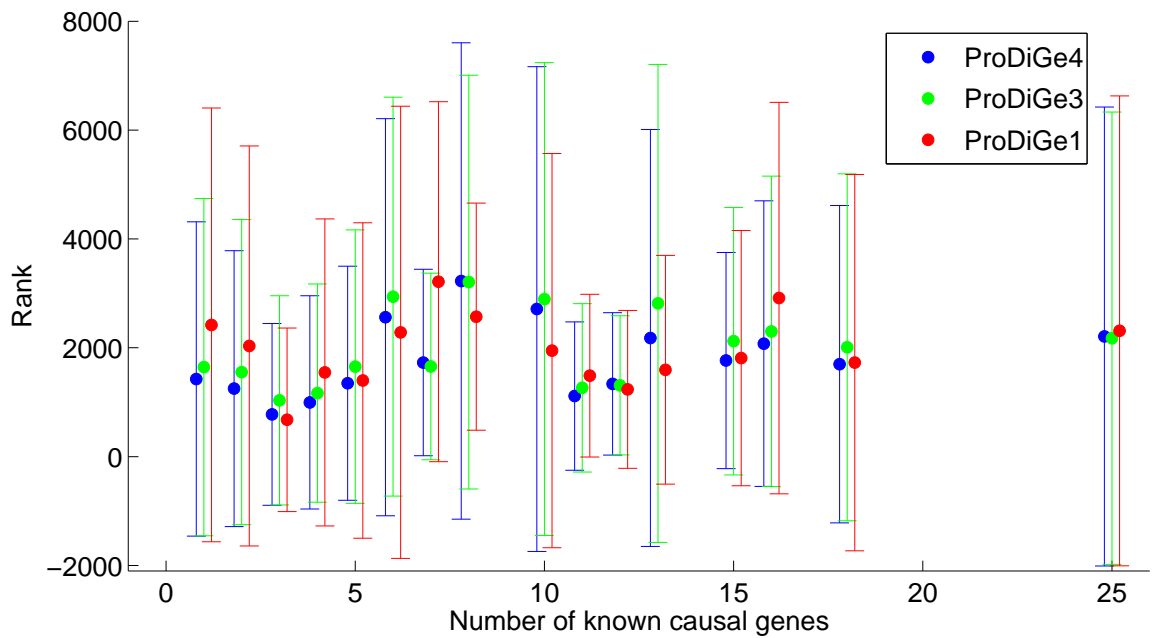}
\end{center}
\caption{
{\bf Effect of the number of related genes on the performance.} 
}
\label{fig:nbgenes}
\end{figure}

Finally, except for the work of \cite{Calvo2007partially}, the PU learning point of view on this long-studied gene prioritization problem is novel. Classical one-class approaches which learn a scoring function to rank candidate genes using known disease genes only are prone to over-fitting in large dimensions when the training set if small, which results in poor performance. We observed that our PU learning strategy, augmented by a multitask point of view to share information across diseases, was useful to obtain better results in the disease gene identification task. In fact, learning from positive and unlabeled examples is a common situation in bioinformatics, and PU learning methods combined or not with multitask kernels have a good potential for solving many problems such as pathway completion, prioritization of cancer patients with a higher risk of relapse, or prediction of protein-protein or protein-ligand interactions.

\section*{Material and Methods}
\label{section:methods}

\subsection*{The gene prioritization problem}

Let us first formally define the disease gene prioritization problem we aim to solve. We start from a list of N human genes $\mathcal{G}=\{G_1,\ldots, G_N\}$, which typically can be the full human genome or a subset of interest where disease genes are suspected. A multitude of data sources to characterize these genes are given, including for instance expression profiles, functional annotation, sequence properties, regulatory information, interactions, literature data, etc... We assume that for each data source, each gene $G\in\mathcal{G}$ is represented by a finite- or infinite-dimensional vector $\Phi(G)$, which defines an inner product $K(G,G') = \Phi(G)^\top \Phi(G')$ between any two genes $G$ and $G'$. $K$ is called a \emph{kernel} in the machine learning community \cite{Scholkopf2002Learning}. Intuitively, $K(G,G')$ may be thought of as a measure of similarity between genes $G$ and $G'$ according to the representation defined by $\Phi$. Since several representations are available, we assume that $L$ feature vector mappings $\Phi_1,\ldots, \Phi_L$ are available, corresponding to $L$ kernels for genes $K_1, K_2, \ldots, K_L$.
Finally, we suppose given a collection of $M$ disorders or disease phenotypes $\mathcal{D}=\{D_1,\ldots, D_M\}$. For each disorder $D_i$, the learner is given a  set of genes $P_i \subset \mathcal{G}$, which contains known causal genes for this phenotype, and a set of candidate genes $U_i \subset \mathcal{G}$ where we want to find new disease genes for $D_i$. Typically $U_i$ can be the complement set of $P_i$ in $\mathcal{G}$ if no further information about the disease is available, or could be a smaller subset if a short list of candidate genes is given for the disease $D_i$. For each disease $D_i$, our goal is to retrieve more causal genes for $D_i$ in $U_i$. In practice, we aim at ranking the elements of $U_i$ from the most likely disease gene to the less likely, and we assess the quality of a ranking by its capacity to rank the true disease genes at or near the top of the list.

\subsection*{Gene prioritization for a single disease and a single data source}
\label{sec:ProDiGe1}
Let us first describe our gene prioritization approach ProDiGe for a single disease ($M=1$) and a single data source ($L=1$). In that case, we are given a single list of disease genes $P\subset \mathcal{G}$, and must rank the candidate genes in $U\subset \mathcal{G}$ using the kernel $K$. As explained in the Introduction, most existing approaches define a scoring function $s:U\rightarrow\mathbb{R}$, using only positive examples in $P$, to quantify how similar a gene $G$ in $U$ is to the known disease genes in $P$. Here we propose to learn the scoring function $s(.)$ both from $P$ and $U$, by formulating the problem as an instance of PU learning.\\

Intuitively, the motivation behind PU learning is to exploit the information provided by the distribution of unlabeled examples to improve the scoring function, as illustrated in Figure 1
. Here we initially have a set of positive examples (genes known to be related to a given disease for instance) which are represented on the graph by blue crosses, and we want to retrieve more of them. Traditional approaches which define a scoring function from $P$ usually try to estimate the support of the positive class distribution to define an area of ``similar genes'', which could be in that case delimited by the dashed line. Now suppose that we additionally observe a set of unlabeled examples (candidate genes), represented by U letters. Green Us are positive unlabeled and red ones are negative unlabeled, but this information is not available. Then, we can have the feeling that the boundary should rather be set in the low density area as represented by the solid line, which better captures reality than the dashed line. Consequently, using the distribution of $U$ in addition to the positive examples can help us better characterize the area of positive examples. This is particularly true in high dimension with few examples, where density estimation from a few positive examples is known to be very challenging.\\

\begin{figure}[!ht]
\begin{center}
\includegraphics[width=0.5\textwidth]{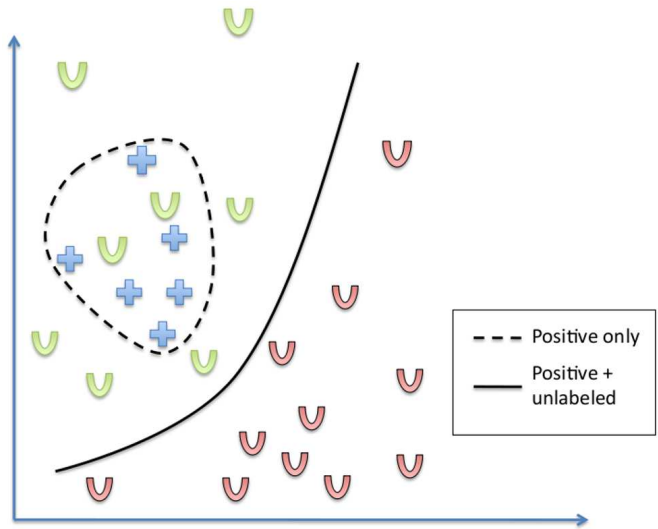}
\end{center}
\caption{
{\bf An intuitive example of how the unlabeled examples could be helpful.}
}
\label{fig:intuitionPUL}
\end{figure}

In practice, a simple and efficient strategy to solve a PU learning problem is to assign negative labels to elements in $U$, and train a binary classifier to discriminate $P$ from $U$, allowing errors in the training labels. Assuming that the binary classifier assigns a score to each point during training (which is the case of, e.g., logistic regression or SVM), the score of an element in $U$ is then just the scored assigned to it by the classifier after training. This approach is easy to implement and it has been shown that building a classifier that discriminates the positive from the unlabeled set is a good proxy to building a classifier that discriminates the positive from the negative set. When the binary classifier used is a SVM, this approach leads to the biased SVM of \cite{Liu2002Partially}, which was recently combined with bagging to reach faster training time and equal performance \cite{Mordelet2010bagging}. In practice, the biased SVM over-weights positive examples during training to account for the fact that they represent high-confidence examples whereas the ``negative'' examples are known to contain false negatives, namely, those we hope to discover. Here we use the variant of \cite{Mordelet2010bagging}, which adds a bootstrap procedure to biased SVM. The additional bagging-like feature takes advantage of the contaminated nature of the unlabeled set, allowing to reach the same performances while increasing both speed and scalability to large datasets. The algorithm takes as input a positive and an unlabeled set of examples, and a parameter $B$ specifying the number of bootstrap iterations. It discriminates the positive set from random subsamples of the unlabeled set and aggregates the successive classifiers into a single one (bootstrap aggregating). The output is a score function $s$ such that for any example $G$, $s(G)$ reflects our confidence that $G$ is a positive example. We then rank elements in $U$ by decreasing score. For more details on the method, we refer the reader to \cite{Mordelet2010bagging}. In practice, we implement the SVM with the libsvm implementation \cite{Chang2001LIBSVM}. After observing in preliminary experiments that the regularization parameter $C$ of the SVM did not dramatically affect the final performance, we set it constant to the default value $C=1$ for all results shown below. The number of bootstrap iterations was set to $B=30$.\\

\subsection*{Gene prioritization for a single disease and multiple data sources}
\label{sec:ProDiGemkl}
When several data sources are available to characterize genes, e.g., gene expression profiles and sequence features, we extend our PU learning method to learn simultaneously from multiple heterogeneous sources of data through \emph{kernel data fusion} \cite{Lanckriet2004statistical}. Formally, each data source is encoded in a kernel function, resulting in $L\geq 1$ kernels $K_1,\ldots,K_L$. We investigate the following two strategies to fuse the $L$ data sources.

First, we simply define a new kernel which integrates the information contained in all kernels as the mean of the $L$ kernels, i.e., we define:
\begin{equation}\label{eq:kint}
K_{int} = \frac{1}{L}\sum_{i=1}^L K_i\,.
\end{equation}
In other words, the kernel similarity $K_{int}(G,G')$ between two genes is defined as the mean similarity between the two genes over the different data sources. This simple approach is widely used and often leads to very good performance with SVM to learn classification models from heterogeneous information \cite{Pavlidis2002Learning,Yamanishi2004Protein,Bleakley2007Supervised}. In our setting, we simply use the integrated kernel \eqref{eq:kint} each time a SVM is trained in the PU learning algorithm described in Section \ref{sec:ProDiGe1}, to estimate a prioritization score from multiple data sources.

Alternatively, we test a method for \emph{multiple kernel learning (MKL)} proposed by \cite{Lanckriet2004Learning,Lanckriet2004statistical}, which amounts to building a weighted convex combination of kernels of the form
\begin{equation}\label{eq:mkl}
K_{MKL} = \frac{1}{L}\sum_{i=1}^L \beta_i K_i\,,
\end{equation}
where the non-negative weights $\beta_i$ are automatically optimized during the learning phase of a SVM. By weighting differently the various information sources, the MKL formulation can potentially discard irrelevant sources or give more importance to gene descriptors with more predictive power. Again, combining MKL with our PU learning strategy described in Section \ref{sec:ProDiGe1} is straightforward: we simply use the MKL formulation of SVM instead of the classical SVM each time a SVM is trained.

\subsection*{Gene prioritization for multiple diseases and multiple data sources}
\label{sec:ProDiGe2}
When multiple diseases are considered, a first option is to treat the diseases independently from each other, and apply the gene prioritization strategy presented in Sections \ref{sec:ProDiGe1} and \ref{sec:ProDiGemkl} to each disease in turn. However, it is known that disease genes share some common characteristics \cite{Lopez-Bigas2004Genome-wide,Adie2005Speeding,Calvo2007partially}, and that similar diseases are often caused by similar genes \cite{Freudenberg2002similarity-based,Ala2008Prediction,Wu2008Network-based,Kohler2008Walking,Vanunu2010Associating,Hwang2010Heterogeneous}. This suggests that, instead of treating each disease separately, it may be beneficial to consider them jointly and share information of known disease genes across diseases. By mutualizing information across diseases, one may in particular attempt to prioritize genes for orphan diseases, with no known causal gene. This is an important property since these diseases are obviously those for which predictions are the most needed.\\

We propose to jointly solve the gene prioritization problem for different diseases by formulating it as a \emph{multitask} learning problem, and we adapt the multitask learning strategy of \cite{Evgeniou2005Learning} to our PU learning framework. In this setting, instead of just learning a scoring function over individual genes $G\in\mathcal{G}$ to rank candidates for a disease, we learn a scoring function over \emph{disease-gene pairs} of the form $(D,G)\in\mathcal{D}\times\mathcal{G}$. In order to learn this scoring function, instead of starting from a set of positive examples $P\subset\mathcal{G}$ made of known disease genes for a particular disease, we start from a set of positive pairs $\big(D_{d(i)},G_{g(i)}\big)_{i=1,\ldots,T}\subset\mathcal{D}\times\mathcal{G}$ obtained by combining the pairs where gene $G_{g(i)}$ is known to be a disease gene for disease $D_{d(i)}$. $T$ is then the total number of known disease-gene pairs. Given the training set of disease-gene pairs, we then learn a scoring function $s$ over $\mathcal{D}\times\mathcal{G}$ using our general PU learning algorithm described in Section \ref{sec:ProDiGe1}, where the kernel function between two disease-gene pairs $(D,G)$ and $(D',G')$ is of the form:
\begin{equation}
\label{eq:pairkernel}
K_{pair}\big( (D,G), (D',G')\big) = K_{disease}(D,D') \times K_{gene}(G,G')\,.
\end{equation}
In this equation, $K_{gene}$ is a kernel between genes, typically equal to one of the kernels described in Sections \ref{sec:ProDiGe1} and \ref{sec:ProDiGemkl} in the context of gene prioritization for a single disease. $K_{disease}$ is a kernel between diseases, which allows sharing of information across diseases, as in classical multitask learning with kernels \cite{Evgeniou2005Learning,Jacob2008Efficient,Jacob2008Protein}. More precisely, we consider the following variants for $K_{pair}$, which give rise to various gene prioritization methods:
\begin{itemize}
\item The \emph{Dirac kernel} is defined as
\begin{equation}\label{eq:dirackernel}
K_{Dirac}(D,D') = 
\begin{cases}
1 &\text{if }D=D',\\
0 &\text{otherwise.}
\end{cases}
\end{equation}
Plugging the Dirac kernel into (\ref{eq:pairkernel}), we see that the pairwise kernel between two disease-gene pairs for different diseases is $0$. One can then show that there is no sharing of information across diseases, and that learning over pairs in this context boils down to treating each disease independently from the others \cite{Evgeniou2005Learning,Jacob2008Efficient,Jacob2008Protein}. This is thus our baseline strategy, which treats each disease in turn, and does not provide a solution for orphan diseases. We call this method ProDiGe1 below.
\item The \emph{multitask kernel} is defined by
\begin{equation}\label{eq:mtkernel}
K_{multitask}(D,D') = 1 + K_{Dirac}(D,D')\,.
\end{equation}
This kernel, which was proposed by \cite{Evgeniou2005Learning}, allows a basic sharing of information across diseases. In addition to the genes known to be causal for a disease of interest through the Dirac kernel, the addition of a constant in \eqref{eq:mtkernel} allows all other known disease genes for other diseases to play the role of positive training examples, although to a lesser extent than the disease genes for the disease of interest. Here we do not use any specific knowledge about the different diseases and their similarity, and simply try to capture properties that may be shared by disease genes in general. This corresponds to a low information prior because a disease equally exploits knowledge about all other diseases. We call this variant ProDiGe2 below.
\item The \emph{phenotype} kernel is an attempt to capture phenotypic similarities between diseases to control the sharing of information across diseases. Indeed, many previous works have used as prior knowledge the fact that similar phenotypes are likely to be caused by similar genes \cite{Freudenberg2002similarity-based,Ala2008Prediction,Kohler2008Walking,Lage2007,Wu2008Network-based,Vanunu2010Associating,Hwang2010Heterogeneous}. This suggests that, instead of sharing information uniformly across diseases as the multitask kernel \eqref{eq:mtkernel} does, it may be beneficial to do it in a more principled way. In particular, the more similar two diseases are, the more they should share information. In practice, this is obtained by defining a kernel $K_{phenotype}$ between diseases that measures their phenotypic similarity, and plugging it into the general pairwise kernel \eqref{eq:pairkernel}. Here we propose to use the phenotypic similarity measure for diseases based on text mining proposed by \cite{Driel2006text-mining}. Since this measure is derived as a correlation measure, the matrix whose entries contain the pairwise similarity measures is eligible for kernel learning. We call the resulting gene prioritization method ProDiGe3 below.
\item The \emph{phenotype+Dirac} kernel. Finally, we propose a slight variant to the phenotype kernel by adding to it the Dirac kernel:
\begin{equation}\label{eq:kernelPD}
K_{P+D}(D,D') = K_{phenotype}(D,D') + K_{Dirac}(D,D')\,.
\end{equation}
The motivation for this kernel is that, since the description of disease phenotypes we use to build $K_{phenotype}$ is incomplete and does not fully characterize the disease, it may occur that two different diseases, with different disease genes, have similar or even identical phenotypic description. In this case, the addition of the Dirac kernel in \eqref{eq:kernelPD} allows to still distinguish  different diseases, and give more importance to the genes associated to the disease of interest than to the genes associated to different diseases with similar phenotypes. We call ProDiGe4 the resulting gene prioritization method.
\end{itemize}
In summary, each of the four kernels for diseases presented above can be plugged into (\ref{eq:pairkernel}) to define a kernel for disease-gene pairs. Then, the PU learning strategy presented in Section \ref{sec:ProDiGe1} can be applied to learn a scoring function over $\mathcal{D}\times\mathcal{G}$. Finally, the ranking of candidate genes in $U_i$ for a particular disease $D_i$ is obtained by decreasing score $s(D_i,G)$ for $G\in U_i$. We thus obtain four variants summarized in Table \ref{tab:prodigevariants}.\\

\begin{table}[ht!]
\begin{center}
\begin{tabular}{|c|c|l|}
\hline
Name & Disease kernel & Sharing of disease gene information across diseases\\
\hline
Prodige1 & $K_{Dirac}$ & No sharing.\\
Prodige2 & $1+K_{Dirac}$ & Uniform sharing. \\
Prodige3 & $K_{phenotype}$ & Sharing weighted by phenotypic similarity.\\
Prodige4 & $K_{Dirac}+K_{phenotype}$ & Sharing weighted by phenotypic similarity and disease identity.\\
\hline
\end{tabular}
\end{center}
\caption{{\bf Summary of ProDiGe variants.}
 We propose four variants, which differ in the way they share information across diseases, as summarized in the third column of the table. The second column shows the kernel for diseases used by each variant to achieve the sharing of information. Apart from the choice of disease kernel, the four variants follow exactly the same procedure described in Section \ref{sec:ProDiGe2}.
}
\label{tab:prodigevariants}
\end{table}%

When heterogeneous sources of information for genes are available, the two strategies proposed in Section \ref{sec:ProDiGemkl} can be easily combined with each of the four ProDiGe variants, since each particular gene kernel translates into a particular disease-gene kernel through \eqref{eq:pairkernel}. In the experiments below, we only implement the MKL approach for ProDiGe1 to compare it to the mean kernel strategy. For other variants of ProDiGe, we restrict ourselves to the simplest strategy where the different information sources are fused through kernel averaging.\\

\subsection*{Experimental setting}

We assess the performance of various gene prioritization methods by leave-one-out cross-validation (LOOCV) on the dataset of known disease-gene association extracted from the OMIM database. Given the list of all disease-gene associations $(D_{d(i)},G_{g(i)})_{i=1,\ldots,T}$ in OMIM, we remove each pair $(D_{d(i)},G_{g(i)})$ in turn from the training set, train the scoring function from the $T-1$ remaining positive pairs, rank all genes $G$ not associated to $D_{d(i)}$ in the training set by decreasing score $s(D_{d(i)},G)$, and check how well $G_{g(i)}$ is ranked in the list. Note that in this setting, we implicitly assume that the candidate genes for a disease are all genes not known to be associated to the disease, i.e., $U_i = \mathcal{G} \backslash P_i$. In the LOOCV setting, each time a pair $(D_{d(i)},G_{g(i)})$ is removed from the training set, the ranking is then performed on $U_{d(i)} \cup \{G_{g(i)}\}$. We monitor the success of the prioritization by the rank of $G_{g(i)}$ among candidate genes in $U_{d(i)}$. Since we are doing a LOOCV procedure, the rank of the left-out sample is directly related to the classical area under the Receiver Operating Characteristics curve (AUC), via the formula $rank = (|U| + 1) \times (1-AUC)$. Therefore, an easy way to visualize the performance of a gene prioritization method is to plot the empirical cumulative distribution function (CDF) of the ranks obtained for all associations in the training set in the LOOCV procedure. For a given value of the rank $k$, the CDF at level $k$ is defined as the proportion of associations $D_{d(i)},G_{g(i)}$ for which gene $G_{g(i)}$ ranked among the top $k$ in the prioritization list for disease $D_{d(i)}$, which can also be called the \emph{recall} as a function of $k$.\\

\subsection*{Other gene prioritization methods}
We compare ProDiGe to two state-of-the-art gene prioritization methods. First we consider the 1-SVM L2-MKL from \cite{Yu2010L2-norm}, which extends and outperforms the Endeavour method \cite{Yu2010L2-norm}, and which we denote MKL1class below. This method performs one-class SVM \cite{Scholkopf2001Estimating} while optimizing the linear combination of gene kernels with a MKL approach in the same time. We downloaded a Matlab implementation of all functions from the supplementary information website of \cite{Yu2010L2-norm}. We used as input the same 9 kernels as for ProDiGe, and we set the regularization parameter of the algorithm $\nu=0.5$, as done by \cite{Yu2010L2-norm}. Second, we compare ProDiGe to the PRINCE method introduced by \cite{Vanunu2010Associating}, which is designed to share information across the diseases. Prior information consists in gene labels that are a function of their relatedness to the query disease. They are higher for genes known to be directly related to the query disease, high but at a lesser extent for genes related to a disease which is very similar to the query, smaller for genes related to a disease that bears little similarity to the query and zero for genes not related to any disease. PRINCE propagates these labels on a PPI network and produces gene scores that vary smoothly over the network. We used the same PPI network for PRINCE as the one used by ProDiGe.

\subsection*{Data}
The first type of data required by ProDiGe is the description of the set $\mathcal{G}$ of human genes. We borrowed the dataset of \cite{Bie2007Kernel-based}, based on Ensembl v39 and which contains multiple data sources. We removed genes whose ID had a ``retired'' status in Ensembl v59, leaving us with 19540 genes. These genes are described by microarray expression profiles from \cite{Son2005Database} and \cite{Su2002Large-scale} (MA1, MA2), expressed sequence tag data (EST), functional annotation (GO) , pathway membership (KEGG), protein-protein interactions from the Human Protein Reference Database (PPI), transcriptional motifs (MOTIF), protein domain activity from InterPro (IPR) and literature data (TEXT). For PPI data which consists in a graph of interactions, a diffusion kernel with parameter 1 was computed to obtain a kernel for genes \cite{Kondor2002Diffusion}. All other data sources provide a vectorial representation of a gene. The inner product between these vectors defines the kernel we create from each data source. All kernels are normalized to unit diagonal to ensure that kernel values are comparable between different data sources, using the formula:
 \begin{equation}
 \label{eq:norm}
 \tilde{K}(G,G') \leftarrow \frac{K(G,G')}{\sqrt{K(G,G)\times K(G',G')}}\,.
 \end{equation}

Second, to define the phenotype kernel between diseases we borrow the phenotypic similarity measure of \cite{Driel2006text-mining}. The measure they propose is obtained by automatic text mining. A disease is described in the OMIM database by a text record \cite{McKusick2007Mendelian}. In particular, its description contains terms from the Mesh (medical subject headings) vocabulary. \cite{Driel2006text-mining} assess the similarity between two diseases by comparing the Mesh terms content of their respective record in OMIM. We downloaded the similarity matrix for 5080 diseases from the MimMiner webpage.

Finally, we collected disease-gene associations from the OMIM database \cite{McKusick2007Mendelian}, downloaded on August 8th, 2010. We obtained 3222 disease-gene associations involving 2606 disorders and 2182 genes.

\section*{Acknowledgments}
We are grateful to L\'eon-Charles Tranchevent, Shi Yu and Yves Moreau for providing the gene datasets, and to Roded Sharan and Oded Magger for making their Matlab implementation of PRINCE available to us. 
This work was supported by ANR grants ANR-07-BLAN-0311-03 and ANR-09-BLAN-0051-04.


\begin{thebibliography}{10}
\providecommand{\url}[1]{\texttt{#1}}
\providecommand{\urlprefix}{URL }
\expandafter\ifx\csname urlstyle\endcsname\relax
  \providecommand{\doi}[1]{doi:\discretionary{}{}{}#1}\else
  \providecommand{\doi}{doi:\discretionary{}{}{}\begingroup
  \urlstyle{rm}\Url}\fi
\providecommand{\bibAnnoteFile}[1]{%
  \IfFileExists{#1}{\begin{quotation}\noindent\textsc{Key:} #1\\
  \textsc{Annotation:}\ \input{#1}\end{quotation}}{}}
\providecommand{\bibAnnote}[2]{%
  \begin{quotation}\noindent\textsc{Key:} #1\\
  \textsc{Annotation:}\ #2\end{quotation}}
\providecommand{\eprint}[2][]{\url{#2}}

\bibitem{Giallourakis2005Disease}
Giallourakis C, Henson C, Reich M, Xie X, Mootha VK (2005) Disease gene
  discovery through integrative genomics.
\newblock Annu Rev Genomics Hum Genet 6: 381--406.
\input{Giallourakis2005Disease}

\bibitem{Perez-Iratxeta2002Association}
Perez-Iratxeta C, Bork P, Andrade MA (2002) Association of genes to genetically
  inherited diseases using data mining.
\newblock Nat Genet 31: 316--319.
\input{Perez-Iratxeta2002Association}

\bibitem{Turner2003POCUS}
Turner FS, Clutterbuck DR, Semple CAM (2003) Pocus: mining genomic sequence
  annotation to predict disease genes.
\newblock Genome Biol 4: R75.
\input{Turner2003POCUS}

\bibitem{Tiffin2005Integration}
Tiffin N, Kelso JF, Powell AR, Pan H, Bajic VB, et~al. (2005) Integration of
  text- and data-mining using ontologies successfully selects disease gene
  candidates.
\newblock Nucleic Acids Res 33: 1544--1552.
\input{Tiffin2005Integration}

\bibitem{Freudenberg2002similarity-based}
Freudenberg J, Propping P (2002) A similarity-based method for genome-wide
  prediction of disease-relevant human genes.
\newblock Bioinformatics 18 Suppl 2: S110--S115.
\input{Freudenberg2002similarity-based}

\bibitem{Aerts2006Gene}
Aerts S, Lambrechts D, Maity S, Van~Loo P, Coessens B, et~al. (2006) Gene
  prioritization through genomic data fusion.
\newblock Nat Biotechnol 24: 537--544.
\input{Aerts2006Gene}

\bibitem{Bie2007Kernel-based}
De~Bie T, Tranchevent LC, van Oeffelen LMM, Moreau Y (2007) Kernel-based data
  fusion for gene prioritization.
\newblock Bioinformatics 23: i125--i132.
\input{Bie2007Kernel-based}

\bibitem{Linghu2009Genome-wide}
Linghu B, Snitkin E, Hu Z, Xia Y, Delisi C (2009) Genome-wide prioritization of
  disease genes and identification of disease-disease associations from an
  integrated human functional linkage network.
\newblock Genome Biol 10: R91.
\input{Linghu2009Genome-wide}

\bibitem{Hwang2010Heterogeneous}
Hwang T, Kuang R (2010) A heterogeneous label propagation algorithm for disease
  gene discovery.
\newblock In: Proceedings of the SIAM International Conference on Data Mining,
  SDM 2010, April 29 - May 1, 2010, Columbus, Ohio, USA. pp. 583--594.
\input{Hwang2010Heterogeneous}

\bibitem{Yu2010L2-norm}
Yu S, Falck T, Daemen A, Tranchevent LC, Suykens Y, et~al. (2010) L2-norm
  multiple kernel learning and its application to biomedical data fusion.
\newblock BMC Bioinformatics 11: 309.
\input{Yu2010L2-norm}

\bibitem{Ala2008Prediction}
Ala U, Piro R, Grassi E, Damasco C, Silengo L, et~al. (2008) Prediction of
  human disease genes by human-mouse conserved coexpression analysis.
\newblock PLoS Comput Biol 4: e1000043.
\input{Ala2008Prediction}

\bibitem{Wu2008Network-based}
Wu X, Jiang R, Zhang M, Li S (2008) Network-based global inference of human
  disease genes.
\newblock Mol Syst Biol 4: 189.
\input{Wu2008Network-based}

\bibitem{Kohler2008Walking}
K{\"o}hler S, Bauer S, Horn D, Robinson P (2008) Walking the interactome for
  prioritization of candidate disease genes.
\newblock Am J Hum Genet 82: 949--958.
\input{Kohler2008Walking}

\bibitem{Vanunu2010Associating}
Vanunu O, Magger O, Ruppin E, Shlomi T, Sharan R (2010) Associating genes and
  protein complexes with disease via network propagation.
\newblock PLoS Comput Biol 6: e1000641.
\input{Vanunu2010Associating}

\bibitem{Tranchevent2010guide}
Tranchevent LC, Capdevila FB, Nitsch D, De~Moor B, De~Causmaecker P, et~al.
  (2010) A guide to web tools to prioritize candidate genes.
\newblock Brief Bioinform 11.
\input{Tranchevent2010guide}

\bibitem{Liu2002Partially}
Liu B, Lee WS, Yu PS, Li X (2002) Partially supervised classification of text
  documents.
\newblock In: ICML '02: Proceedings of the Nineteenth International Conference
  on Machine Learning. San Francisco, CA, USA: Morgan Kaufmann Publishers Inc.,
  pp. 387--394.
\input{Liu2002Partially}

\bibitem{Denis2005Learning}
Denis F, Gilleron R, Letouzey F (2005) Learning from positive and unlabeled
  examples.
\newblock Theor Comput Sci 348: 70--83.
\input{Denis2005Learning}

\bibitem{Mordelet2010bagging}
Mordelet F, Vert JP (2010) A bagging {SVM} to learn from positive and unlabeled
  examples.
\newblock Technical Report HAL-00523336.
\input{Mordelet2010bagging}

\bibitem{Evgeniou2005Learning}
Evgeniou T, Micchelli C, Pontil M (2005) Learning multiple tasks with kernel
  methods.
\newblock J Mach Learn Res 6: 615-637.
\input{Evgeniou2005Learning}

\bibitem{Jacob2008Efficient}
Jacob L, Vert JP (2008) Efficient peptide-{MHC}-{I} binding prediction for
  alleles with few known binders.
\newblock Bioinformatics 24: 358--366.
\input{Jacob2008Efficient}

\bibitem{Jacob2008Protein}
Jacob L, Vert JP (2008) Protein-ligand interaction prediction: an improved
  chemogenomics approach.
\newblock Bioinformatics 24: 2149--2156.
\input{Jacob2008Protein}

\bibitem{Pavlidis2002Learning}
Pavlidis P, Weston J, Cai J, Noble W (2002) Learning gene functional
  classifications from multiple data types.
\newblock J Comput Biol 9: 401--411.
\input{Pavlidis2002Learning}

\bibitem{Schoelkopf2004Kernel}
Sch{\"o}lkopf B, Tsuda K, Vert JP (2004) Kernel {M}ethods in {C}omputational
  {B}iology.
\newblock The MIT Press, Cambridge, Massachussetts: MIT Press.
\input{Schoelkopf2004Kernel}

\bibitem{Lanckriet2004statistical}
Lanckriet GRG, De~Bie T, Cristianini N, Jordan MI, Noble WS (2004) A
  statistical framework for genomic data fusion.
\newblock Bioinformatics 20: 2626-2635.
\input{Lanckriet2004statistical}

\bibitem{McKusick2007Mendelian}
McKusick V (2007) Mendelian inheritance in man and its online version, omim.
\newblock Am J Hum Genet 80: 588--604.
\input{McKusick2007Mendelian}

\bibitem{Brancotte2011Gene}
Brancotte B, Biton A, Bernard-Pierrot I, Radvanyi F, Reyal F, Cohen-Boulakia
  S (2011) Gene list significance at-a-glance with GeneValorization.
\newblock Bioinformatics 27: 1187--1189.
\input{Brancotte2011Gene}

\bibitem{Calvo2007partially}
Calvo B, L{\'o}pez-Bigas N, Furney S, Larra{\~n}aga P, Lozano J (2007) A
  partially supervised classification approach to dominant and recessive human
  disease gene prediction.
\newblock Comput Methods Programs Biomed 85: 229--237.
\input{Calvo2007partially}

\bibitem{Scholkopf2002Learning}
Sch{\"o}lkopf B, Smola AJ (2002) Learning with {K}ernels: {S}upport {V}ector
  {M}achines, {R}egularization, {O}ptimization, and {B}eyond.
\newblock Cambridge, MA: MIT Press.
\input{Scholkopf2002Learning}

\bibitem{Chang2001LIBSVM}
Chang CC, Lin CJ (2001) {LIBSVM}: a library for support vector machines.
\newblock Software available at \url{http://www.csie.ntu.edu.tw/~cjlin/libsvm}.
\input{Chang2001LIBSVM}

\bibitem{Yamanishi2004Protein}
Yamanishi Y, Vert JP, Kanehisa M (2004) Protein network inference from multiple
  genomic data: a supervised approach.
\newblock Bioinformatics 20: i363-i370.
\input{Yamanishi2004Protein}

\bibitem{Bleakley2007Supervised}
Bleakley K, Biau G, Vert JP (2007) Supervised reconstruction of biological
  networks with local models.
\newblock Bioinformatics 23: i57--i65.
\input{Bleakley2007Supervised}

\bibitem{Lanckriet2004Learning}
Lanckriet G, Cristianini N, Bartlett P, El~Ghaoui L, Jordan M (2004) Learning
  the kernel matrix with semidefinite programming.
\newblock J Mach Learn Res 5: 27-72.
\input{Lanckriet2004Learning}

\bibitem{Lopez-Bigas2004Genome-wide}
L{\'o}pez-Bigas N, Ouzounis CA (2004) Genome-wide identification of genes
  likely to be involved in human genetic disease.
\newblock Nucleic Acids Res 32: 3108--3114.
\input{Lopez-Bigas2004Genome-wide}

\bibitem{Adie2005Speeding}
Adie EA, Adams RR, Evans KL, Porteous DJ, Pickard BS (2005) Speeding disease
  gene discovery by sequence based candidate prioritization.
\newblock BMC Bioinformatics 6: 55.
\input{Adie2005Speeding}

\bibitem{Lage2007}
Lage K, Karlberg E, Størling Z, Olason P, Pedersen A, et~al. (2007) A human
  phenome-interactome network of protein complexes implicated in genetic
  disorders.
\newblock Nat Biotechnol 25: 309--316.
\input{Lage2007}

\bibitem{Driel2006text-mining}
van Driel M, Bruggeman J, Vriend G, Brunner H, Leunissen J (2006) A text-mining
  analysis of the human phenome.
\newblock Eur J Hum Genet 14: 535--542.
\input{Driel2006text-mining}

\bibitem{Scholkopf2001Estimating}
Sch{\"o}lkopf B, Platt JC, Shawe-Taylor J, Smola AJ, Williamson RC (2001)
  Estimating the support of a high-himensional distributions.
\newblock Neural Comput 13: 1443--1471.
\input{Scholkopf2001Estimating}

\bibitem{Son2005Database}
Son C, Bilke S, Davis S, Greer B, Wei J, et~al. (2005) Database of m{RNA} gene
  expression profiles of multiple human organs.
\newblock Genome Res 15: 443--450.
\input{Son2005Database}

\bibitem{Su2002Large-scale}
Su A, Cooke M, Ching K, Hakak Y, Walker J, et~al. (2002) Large-scale analysis
  of the human and mouse transcriptomes.
\newblock Proc Natl Acad Sci U S A 99: 4465--4470.
\input{Su2002Large-scale}

\bibitem{Kondor2002Diffusion}
Kondor RI, Lafferty J (2002) Diffusion kernels on graphs and other discrete
  input.
\newblock In: Proceedings of the Nineteenth International Conference on Machine
  Learning. San Francisco, CA, USA: Morgan Kaufmann Publishers Inc., pp.
  315--322.
\input{Kondor2002Diffusion}

\end{thebibliography}

\end{document}